\documentclass[12pt,preprint]{aastex}
\usepackage{natbib}
\usepackage{amsmath,amsfonts,amssymb}
\usepackage{graphicx}
\def\idm#1{{\mbox{\scriptsize #1}}}

\newcommand\Ym{\langle Y\rangle}

\def\deg{{\rm o}}
\def\idm#1{{\mbox{\scriptsize #1}}}

\def\astrobj#1{#1\ }

\def\url#1{\texttt{#1}}
\newcommand\pstar{{\astrobj{HD~160691}}}
\newcommand\Chi{{(\chi^2_\nu)^{1/2}}}
\newcommand\Chim{{(\chi^2_\nu)^{1/2}_{\idm{max}}}}
\newcommand\ms{\mbox{ms}^{-1}}
\newcommand\bl{{\mathbf l}} 
\newcommand\bm{{\mathbf m}}

\newcommand\br{{\mathbf r}}

\renewcommand\bv{{\mathbf v}}

\newcommand\bP{{\mathbf P}}
\newcommand\bQ{{\mathbf Q}} 
\newcommand\bR{{\mathbf R}}

\newcommand\bV{{\mathbf V}}

\newcommand\brho{\boldsymbol{\rho}}
\newcommand\bupsilon{\boldsymbol{\upsilon}}

\shortauthors{Go\'zdziewski, Konacki, Maciejewski}
\begin{document}
\title{Where is the second planet in the HD~160691 planetary system?}
\author{Krzysztof Go\'zdziewski\altaffilmark{1}}
\affil{Toru\'n Centre for Astronomy, N.~Copernicus University,
Gagarina 11, 87-100 Toru\'n, Poland}
\author{Maciej Konacki\altaffilmark{2}}
\affil{Department of Geological and Planetary Sciences, California
Institute of Technology, MS 150-21, Pasadena, CA 91125, USA \\
Nicolaus Copernicus Astronomical Center, Polish Academy of Sciences,
Rabia\'nska 8, 87-100 Toru\'n, Poland}
\author{Andrzej J. Maciejewski \altaffilmark{3}}
\affil{Institute of Astronomy, University of Zielona G\'ora,
Podg\'orna 50, 65-246 Zielona G\'ora, Poland}
\altaffiltext{1}{e-mail: k.gozdziewski@astri.uni.torun.pl}
\altaffiltext{1}{e-mail: maciej@gps.caltech.edu}
\altaffiltext{3}{e-mail: maciejka@astro.ia.uz.zgora.pl}
\begin{abstract}
The set of radial velocity measurements of the \pstar has been recently
published by  \cite{Jones2003}. It reveals a linear trend that indicates a
presence of the second planet in this system. The  preliminary
double-Keplerian orbital fit to the observations, announced by the discovery
team, describes a highly unstable, self-disrupting configuration. Because the
observational window of the \pstar system is narrow, the orbital parameters
of the hypothetical second companion are unconstrained. In this paper we try
to find out whether a second giant  planet can exist up to the distance of
Jupiter and search for the dynamical constraints on its orbital parameters.
Our analysis employs a combination of fitting algorithms and  simultaneous
examination of the dynamical stability of the obtained orbital fits. It
reveals that if the semi-major axis of the second planet is smaller than
$\simeq 5.2$~AU,  the observations are consistent with  quasi-periodic,
regular motions of the system confined to the islands of various low-order
mean motion resonances, e.g., 3:1, 7:2, 4:1, 5:1, or to their vicinity. In
such cases the second planet has smaller eccentricity $\simeq 0.2-0.5$ than
estimated in the previous works. We show that the currently available Doppler
data   rather preclude the 2:1 mean motion resonance expected by some authors
to be present in the \pstar system. We also demonstrate that the
MEGNO-penalty  method (MEGNO is an acronym for the Mean Exponential Growth factor
of Nearby Orbits), developed in this paper, which is a combination of the
genetic minimization  algorithm and the MEGNO stability analysis, can be
efficiently used for predicting  stable planetary configurations when only a
limited number of observations is given  or the data do not provide tight
constraints on the orbital elements.
\end{abstract}
%
%
\section{Introduction}
%
%
Recently, \cite{Jones2003} have published the radial velocity (RV)
observations of a metal-reach solar-type star \pstar. These data, consisting
of 38  RV measurements and having the formal uncertainties in the range
$2.3$--$6.3$~$\ms$,  upgrade the RV data published in \cite{Butler2001} and
in the preprint \citep{Jones2002}, which quotes a smaller number, 33, of the
RV measurements than in its published version. As it will become clear later,
both references are  relevant to our discussion. These new data make it
possible to refine a preliminary initial parameters of the \pstar system
given in \cite{Butler2001}. The new data confirm the Keplerian elements of
the previously detected planet and reveal a linear trend indicating that a
second planet  in the \pstar system is possible. In the preprint
\citep{Jones2002}, the approximate  orbital elements of the second companion
indicate a proximity of the \pstar system to the 2:1 mean motion resonance
(MMR).  Such possibility is puzzling because the \pstar system would be the
third instance of this resonance (in addition to systems around  Gliese~876
and HD~82943) among only a dozen of multi-planetary exosystems known to
date\footnote{http://www.encyclopaedia.fr}.  However, the initial parameters
reproduced in Table~\ref{tab:tab1} as the fit J2, and considered as
osculating Keplerian elements, describe orbits that cross each other. This
leads to a rapid disintegration of the system. The updated orbital fit given
in \cite{Jones2003} is highly unstable, either, and it rather precludes the
inquiring case  of the 2:1 MMR (Table~\ref{tab:tab1}, fit J2a).  However, the
authors stress that in both cases the initial elements of the hypothetical
second planet are not constrained and very uncertain.

In this work we analyze the published RV data of the \pstar  system in a more
general manner --- beyond a formal fit of the Keplerian orbital elements. 
Although the RV data do not provide tight constrains on the orbital elements
of the second planet, these elements are confined by the dynamics, especially
in a case when the second companion is not very far from the inner planet.  
This relevant information is entirely omitted in the Keplerian fit. Applying
the Copernican principle, it is not  likely to observe planetary systems
during moments of their extraordinary orbital evolution \citep{Murray2001}.
But a disruption of a planetary system is an unusual event and we can argue
that if by a formal fit, we find the initial conditions that lead to such a
phenomenon then this fit is unlikely either. In this sense, the system
dynamics becomes  {\em an observable} that should be taken into account
together with the  RV observations while looking for the possible orbital
parameters of a planetary system.  This way of reasoning has already been
applied by skipping orbital fits that lead to highly unstable systems
\citep[see, for instance][]{Stepinski2000,Laughlin2001,
Rivera2001,Gozdziewski2001b}.  It has also inspired us to ask whether the
current,  limited RV data set of the \pstar system merged with the dynamical
analysis of the obtained best fit orbital parameters can provide enough
information to estimate meaningful limits on the orbital parameters of the
hypothetical second planet. In this paper we also describe a method that can
be useful in resolving this  generic problem not only for the \pstar system,
chosen  as an excellent candidate  for such analysis, but in all other cases
when the RV observations do not supply sufficient constraints on the orbital
parameters of planetary companions.
%
%
%
\section{Overall dynamical characteristic of the \pstar system}
%
%
The analysis of the \pstar system in this paper relies on the radial
velocity  observations presented in the preprint by \cite{Jones2002} and
their recently published paper \citep{Jones2003}.  Even simple considerations
lead to the conclusion that the approximate fits published in these papers
are generically unstable for large eccentricities of the putative planet c.
If we assume that $1.5~\mbox{AU} <a_{\idm{c}} <5.2$~AU, then in a stable
system $e_{\idm{c}}$ cannot be larger  than $\simeq 0.6$---otherwise the
orbits  cross each other and a collision occurs. For a reference see
Figure~\ref{fig:fig4}, where we show approximate planetary collision line,
determined through $a_{\idm{b}}(1+e_{\idm{b}}) = a_{\idm{c}}(1-e_{\idm{c}})$,
and calculated for $e_{\idm{b}} = 0.31$ and $a_{\idm{b}}=1.49$~AU.  These
values are justified by the fits of  \cite{Jones2002,Jones2003} and our
solutions. The orbital parameters of the inner planet are well constrained by
the RV data. If $e_{\idm{c}}$ is really large,  larger than the critical
collisional value,  the planets can avoid collisions only when a dynamical
mechanism  prevents them from an encounter. It is well known, that this can
be provided by orbital resonances. These resonances are present in many
multi-planetary  exosysystems \citep[see Table 8 in][]{Fischer2003}. Thus,
assuming that the hypothetical giant planet's orbit is somewhere up to the
Jupiter distance, $\simeq 5.2$~AU from the star, the mutual interactions
between the companions are substantial and it is reasonable to expect the
existence of an orbital resonance in the \pstar system. For the hypothetical
2:1 MMR, it should be accompanied by the secular apsidal resonance (SAR),
$\varpi_{\idm{b}} \simeq \varpi_{\idm{c}}$. This has been studied in detail
in many recent papers about Gliese~876 
\cite[e.g.,][]{Laughlin2001,Rivera2001,Lee2002,Gozdziewski2002,Ji2002,Hadjidemetriou2002}
and HD~82943 \cite[]{Gozdziewski2001b,Hadjidemetriou2002,Jiang2002}.
These systems are striking examples of the protective role of the 2:1 MMR
accompanied by the SAR. These resonances, acting together, permit for stable
planetary configurations even for extremely high eccentricities, $\simeq
0.95-0.98$~\cite[]{Gozdziewski2002}. It could help us explain a very large
eccentricity of the second planet and maintain the system stability in spite
of a generic obscure to a stable orbital evolution which comes from the
possibility of close encounters. In the light of unconstrained orbital
elements of the outer companion, the question whether a stable 2:1 MMR is
permissible in the \pstar system is an attractive  and challenging subject of
study. 

To get an overview of the possible dynamical states of the system we
calculated a number of the MEGNO
\footnote{
The Mean Exponential Growth factor of Nearby Orbits (MEGNO)  is a
technique invented by  \cite{Cincotta2000}. This is so called fast
indicator makes possible a rapid determination whether an
investigated initial condition leads to a quasi-periodic or irregular (chaotic)
motion of a planetary system.  This method is advocated by us for a study
of planetary dynamics in a series of recent papers \citep[see, e.g.,]
[]{Gozdziewski2001a,Gozdziewski2003d}. The  MEGNO integrations in this
paper were driven by  a Bulirsh-Stoer integrator. We used the ODEX code
\cite[]{Hairer1995}. The relative and absolute accuracies of the integrator 
were set to $10^{-14}$ and $5\cdot 10^{-16}$, respectively.
The position of the nominal condition is marked in contour
plots by the intersection of two thin lines. The stable areas are marked in
the MEGNO maps by values close to 2.
}
stability maps in the ($a_{\idm{c}},e_{\idm{c}}$)-plane, centered on the
nominal position of this resonance, about $2.4$~AU. We assumed that the
orbital elements of the inner planet are fixed at the J2 fit  (see
Table~\ref{tab:tab1}). These data have been considered as the osculating,
astrocentric elements at the epoch of the periastron passage of the outer
planet. As we already mentioned, in both the fits J2 and J2a, as well as in
our solutions (they will be described later), the elements of the inner
companion are very similar and we considered them well constrained by the
observations. We assumed that the system is coplanar and edge-on.  This
assumption reduces greatly the number of possible orbital configurations but
the ($a_{\idm{c}},e_{\idm{c}}$)-plane is dynamically representative for the
system dynamics, in the sense that it crosses all resonances
\cite[]{Robutel2001}. Moreover, the widths of MMRs depend on the phases of
companions and for this reason we varied, in subsequent maps, the initial
longitude of periastron and the mean anomaly of the outer planet.  The
results of the experiment are  illustrated in Figure~\ref{fig:fig1}. In the
test, $\omega_{\idm{c}}=99^{\circ}$ is equal to its formal fit value and
corresponds to the first column of the MEGNO-maps which are calculated for
the initial mean anomaly $M_{\idm{c}}=0^{\circ}$, $90^{\circ}$,
$180^{\circ}$  and $270^{\circ}$, respectively, i.e., for different initial
phases of the outer planet.  Simultaneously, the initial orbital phase of the
inner planet has been fixed at $\omega_{\idm{b}}=320^{\circ}$ and
$M_{\idm{b}}\simeq 10^{\circ}$.  The middle  and the right columns in
Figure~\ref{fig:fig1} are for $\omega_{\idm{c}} \simeq 320^{\circ}$ and
$\omega_{\idm{c}} \simeq 120^{\circ}$,  respectively. These values have been
selected in such a way that the apsidal lines are intially 
aligned (Figure~\ref{fig:fig1}, middle column) or antialigned 
(Figure~\ref{fig:fig1}, right column).

The scans corresponding to the nominal initial condition (IC, the first
column in Figure~\ref{fig:fig1}), reveal a number of very narrow zones of
stable motions. They are identified with the low-order MMRs, e.g., 4:3, 7:5,
3:2, 5:3, 2:1, 7:3, 5:2, 8:3 and 3:1.   In the scans for
$M_{\idm{c}}=90^{\circ}$  and  $M_{\idm{c}}=270^{\circ}$,  we discovered
extended islands of stable 2:1~MMR for large $e_{\idm{c}}$.  As we expected,
this resonance is  accompanied by the SAR, with apsidal lines librating about
$180^{\circ}$. This is illustrated in Fig~\ref{fig:fig2}a,b,c.  In the 
MEGNO-scans, we also found a stable island of the 3:1 MMR, which is present
for extremely large eccentricities of the outer planet, up to 0.98! The
evolution of the astrocentric orbital elements is illustrated in
Figure~{\ref{fig:fig2}d,e,f.  Outside the MMRs zones, the planetary
configurations are very unstable and typically  self-disrupt rapidly (on
timescales of a few years). 

As we already mentioned, the IC quoted in \cite{Jones2002} is very close to a
stable island of the 2:1 MMR. In order to lock the dynamics in this
resonance, a change of the orbital phase of the  outer planet from (the
formal value) $M_{\idm{c}} \simeq 0^{\circ}$ to $M_{\idm{c}} \simeq
90^{\circ}$ is required. One would think that such a change may be
permissible since the orbital parameters are weakly constrained by the data,
but in fact, the synthetic RV curve is strongly affected by a modification of
the relative phases of the planets (for an example see section~5).  Actually,
the scans shown in Figure~\ref{fig:fig1}, suggest that   by changing the
weakly constrained or unconstrained parameters of the outer planet, we can
also obtain many other, {\em stable}   resonance configurations. However,
similarly to the previous case, it does not necessarily mean that these
configurations would produce Doppler signals which  are consistent with the
RV observations.  Because the linear trend present in the RV data is
only  at best a "fingerprint" of the second companion, any additional
observations significantly  alter the orbital fit. For these reasons a
detailed, global analysis of the stability of a few orbital configurations of
the \pstar system, defined by the very uncertain fits, is a hopeless task, if
we aspire to investigate systems that are at least similar to the  observed
one. In the next section we show that in fact the linear trend present in the
data permits for a continuum of equally good orbital  fits. This has been
already pointed out by \cite{Jones2003}.

%
%
\section{Global 2-Keplerian fit to the RV data}
%
%
Following the arguments given in in the introduction,  we try to reduce the
number of possible orbital configurations  of the \pstar system by analyzing
$\Chi$-goodness of the fits simultaneously with  the dynamical character of
the resulting IC. In order to find the best double-Keplerian fit, we were
applying the RV models that were subsequently  more and more elaborate. At
first, we assumed that the data can be approximated by a sum of the
one-Keplerian RV signal and a linear trend (the KT model). Note that our
model of the Keplerian RV signal differs from the classical version which is
really suitable for binary stellar systems (see the Appendix for all
details).  To perform a global search for the best fit,  we computed the 
function $\Chi(P_{\idm{b}},e_{\idm{b}})$ in the
$(P_{\idm{b}},e_{\idm{b}})$-plane, fixing $P_{\idm{b}}$ and $e_{\idm{b}}$ in
the ranges $[600-700]$~d and $[0,0.9]$, respectively. The Levenberg-Marquardt
(LM) minimization scheme \citep{Press1992} was used to find the best fit at
every point of the $100\times100$-data grid in this region. We searched for
the best fits at every point of the grid by varying the starting phase of the
inner planet with the step of $30^{\circ}$. The obtained $\Chi$-map reveals a
very  well determined minimum of $\Chi$ (see Fig.~\ref{fig:fig3}a)
corresponding to  $P_{\idm{b}} \simeq 638$~d and $e_{\idm{b}} \simeq 0.31$.
These parameters coincide  very well with the solution given in
\cite{Jones2002} and \cite{Jones2003}. The fit has $\Chi \simeq 1.48$ and rms
$\simeq 4.8 \ms$.

Next, we assumed that the linear trend can be explained as the presence of a
second companion in the system
\footnote{
Although the HD~160691 is unlikely a variable star, we examined its visual
photometry by Hipparcos (http://astro.estec.esa.nl/Hipparcos/). Curiously,
the visual magnitude reveals signs of a linear trend with a small full
amplitude $\simeq 0.04^{\idm{m}}$  and a period $1300$--$1500$~d roughly
coinciding with the approximate period of the outer companion. Nevertheless,
this is highly speculative because the Hipparcos photometry is believed to
be uncertain.
}.
Taking the best fit parameters of the KT model as an approximation  of the 
orbital elements of the inner companion, we calculated the function 
$\Chi(P_{\idm{c}},e_{\idm{c}})$ for the double-Keplerian (JK2) model.  At
every point of the grid,  $P_{\idm{c}} \in [1200,3600]$~d and $e_{\idm{c}}
\in [0,0.9]$, the best fit solution was obtained by varying the initial
longitudes of periastron and the mean anomalies of each companion with the
step of $30^{\circ}$. The result is shown in Fig.~\ref{fig:fig3}b. The best
fit solution [the JK2 fit,  $\Chi \simeq 1.39$ and rms $\simeq 3.9~\ms$, see
Table~\ref{tab:tab2}] corresponds to $P_{\idm{c}} \simeq 1560$~d and
$e_{\idm{c}}\simeq 0.78$. The synthetic RV curve is shown in the left panel
of Figure~\ref{fig:fig5}. In the entire range of $e_{\idm{c}}$, the fits 
obtained for  $P_{\idm{c}} > 1600$~d are of very similar significance because
their $\Chi \simeq 1.5$ and the rms is about $ 4.5~\ms$. This result is not
surprising as it reflects the  limitation caused by a small number of
measurements and a narrow observational window compared to the likely period
of the outer companion.  However, in the range of small $P_{\idm{c}}$,
specifically for the one corresponding to  the 2:1~MMR, the double-Keplerian
fits are significantly worse.

The global fit, obtained through the $(P_{\idm{c}},e_{\idm{c}})$-scan, has
been verified by the minimization method based on the genetic algorithm (GA).
The GAs are still not very popular but they have been proved to be very
useful for finding good starting points for the precise gradient methods of
minimization like e.g. the well known Levenberg-Marquardt scheme. For a very
interesting review of the astronomical applications of the GAs see the paper
by \cite{Charbonneau1995}.  To the best of our knowledge this method of
minimization was used for analyzing the RV data of $\upsilon$~Andr by
\cite{Butler1999} and \cite{Stepinski2000}, the Gliese~876 system by
\cite{Laughlin2001,Laughlin2002a},  and  55~Cnc by \cite{Marcy2002}. We also
applied it to the HD~12661 system \citep{Gozdziewski2003d}. Basically, the
genetic scheme makes it possible to find the {\em global} minimum of the
$\Chi$ function. For computations, we used the publicly available code PIKAIA
ver.
1.2\footnote{http://www.hao.ucar.edu/public/research/si/pikaia/pikaia.html} 
by Paul Charbonneau which allows to limit the search  space. This was very
useful in our situation, because  we could restrict the GA search region to
the one analyzed by the JK2 scan method. In many restarts of the GA code,
confined to  $P_{\idm{c}} \in [1200,3600]$~d, we repeatedly found the best
fit which is given in Table~\ref{tab:tab2} as  the GA2 solution. It is 
qualitatively similar to the best JK2 solution. In an additional set  of the
runs, the range of $P_{\idm{c}}$ has been restricted to [1000,1400]~d to
cover the hypothetical 2:1~MMR. As it turns out,  in this case the best fits
have $\Chi \simeq 1.53$, substantially larger than the GA2 solution and
correspond to $P_{\idm{b}} \simeq 612$~d, $P_{\idm{c}} \simeq 1400$~d. This
is consistent with the results of the JK2 scan search  as the best GA in this
region were found confined to the area around the limiting
$P_{\idm{c}}=1400$~d.

At this point we could say that there is not much more to do unless we try to
analyze the dynamical stability of the obtained  fits. The results of such an
analysis are presented in Figure~\ref{fig:fig4}. Panel~\ref{fig:fig4}a shows
the MEGNO signatures evaluated for every fit in the JK2 scan. For the purpose
of this map the Kepler-Jacobi orbital elements are transformed to the 
osculating, astrocentric elements  at the epoch of the first observation.
This scan reveals that most of the JK2 fits are chaotic and  unstable,
including the best fit solution. (The genetic best fit solution GA2 is 
strongly chaotic too). The border of the dynamical stability, clearly seen 
in all the panels, is well correlated with the planetary collision line
marked with a thick line. The instability, defined through the MEGNO, in the
strict sense of regular and chaotic behaviours, is meaningful and leads to
macroscopic  changes of the orbital elements during a relatively very short
integration time. This is illustrated in Figure~\ref{fig:fig4}b where we
marked the maximal eccentricities of the outer planet attained during the
integration timespan equal to at most $\sim10^4$ orbital periods of this
planet. The integrations were stopped if one of the eccentricities had become
larger than 0.99 or one of the  semi-major axes exceeded 10~AU. In these
cases, the system has been considered as disrupted.  As we could expect,
above the  planetary collision line, the osculating elements are completely
different from their starting values which confirms the dramatic effects of
the close encounters or collisions.

Another interesting result, provided by the analysis of the short term
dynamics, is shown in Figure~{\ref{fig:fig4}c,d}.  It illustrates a detection
of the SAR and an estimate of the semi-amplitude of the critical argument
$\theta =\varpi_{\idm{b}}-\varpi_{\idm{c}}$. It was derived with the same
method which was successfully applied to the HD~12661 system in 
\citep{Gozdziewski2003b}.  In the MEGNO code, we evaluated the  maximal value
$\theta_{\idm{max}}$ of $\theta$, after every step of the renormalization of
the variational equations (the time step was equal to the orbital period of
the outer planet). The maximal value of the critical argument was taken
relative to the center of libration  $0^{\deg}$ or $180^{\circ}$. To avoid 
the effects of a possible transition into the apsidal resonance, the
determination of $\theta_{\idm{max}}$ was started after the first half of the
integration period. Finally, if $\theta_{\idm{max}}<180^{\circ}$, then we
treated this value as a semi-amplitude of the apsidal librations. Although
the period of integrations is relatively very short, such an approximation of
the semi-amplitude already gives much insight into the global dynamics of the
system. Note that in the $\theta$-maps we do not mark the systems which
collided during the integration time (the white areas above the planetary
collision line). Remarkably,  both the $\theta$-maps  reveal a small,
isolated  island about $P_{\idm{c}} \simeq 1900$~d related to the 3:1~MMR. 
In this island the $\theta$ argument  librates about $240^{\circ}$. A much
more extended stable region emerges for $P_{\idm{c}} > 2200$~d. The
configurations in this region correspond  mostly to the librations of
$\theta$ about $180^{\circ}$ (see Figure~{\ref{fig:fig4}d). An obvious and
clear correlation between the MEGNO scan and the MMRs structures is visible
in this plot. These structures  correspond to a number of the low-order MMRs:
e.g., 7:2, 4:1, 5:1.  

Because the best fit solutions JK2 and GA2 are confined to the $e_{\idm{c}}$ 
regions  where the collisions are very likely,  we verified whether the
Jacobi-Keplerian fits properly represent  the dynamical state of the \pstar
system. We compared the synthetic RV curve  obtained from these models and
the RV curve emerging from the full Newtonian mode of the Doppler signal. The
top panels in Figure~\ref{fig:fig5} illustrate the  synthetic RV curves and
the bottom panels are  for the appropriate differences between the RV
signals. For the GA2 solution the difference reaches about 4~$\ms$  over the
time span of the data set and is comparable to the internal accuracy of the
precise RV method. In the case of the JK2 solution, this difference reaches
$\simeq 40~\ms$ and is unacceptably large. This comparison demonstrates the
obvious need of incorporating the mutual interactions between the companions
in the fitted model. This is seen even better in  Figure~{\ref{fig:fig3}c}.
It illustrates the global effects of neglecting the mutual interactions
between the companions in the model. At every point of the JK2-scan
(Figure~\ref{fig:fig3}b) we computed the value of $\Chi$  emerging from the
Keplerian and Newtonian RV signals. Similarly to the previous experiment, the
best JK2 fits have been considered as the osculating elements at the epoch of
the first observation. We plotted $\log |\Delta\Chi|$, where $\Delta\Chi$
corresponds to a difference of $\Chi$ between the models. This figure shows
that the JK2 fits are in fact unacceptable for large $e_{\idm{c}}$. On the
other hand, for $P_{\idm{c}}> 1800$~d and moderately small $e_{\idm{c}}$, 
the Keplerian and Newtonian RV curves coincide perfectly. These results
confirm once again the remarkable result of \cite{Laughlin2001}: for strongly
interacting systems, the Keplerian fitting has a very limited use.

To put this statement into work, we refined the two planet fits using the LM
algorithm driven by the full Newtonian model of the RV signal, and taking the
Keplerian fits as the starting data to the gradient LM search. Similarly to
the previous cases,  these fits have been formally transformed to
astrocentric osculating Keplerian  elements, at the epoch of the first
observation. Because  in the region of large $e_{\idm{c}}$, the JK2 fits do
not describe the system's state properly, they are unlikely optimal starting
points in this area, and it was hard to expect that the LM algorithm will 
find truly global minimum of $\Chi$. To obtain an approximation of the best
Newtonian  $\Chi$-scan and to simplify the search, both the masses and
$a_{\idm{b}}$ have been fixed at their JK2 values.  All other  orbital
parameters have been fitted by the gradient algorithm.  The obtained
$\Chi$-map  in the $(P_{\idm{c}},e_{\idm{c}})$-plane of the starting orbital
elements,  is shown in Figure~\ref{fig:fig3}d. Clearly, for large
$e_{\idm{c}}$  the $N$-body model gives a significant improvement of the
$\Chi$ function over the Keplerian model.  The best Newtonian fit found in
this test is given in Table~\ref{tab:tab2} as the NL2 fit and its synthetic
RV curve is shown in the left panel of Figure~\ref{fig:fig6}. Unfortunately,
this fit is highly unstable too.  Finally, we examined the MEGNO signatures
of the rest of about 20,000 fits, with about 17,300 cases when $\Chi<1.5$.
Only about 260 initial conditions appeared to be dynamically stable, having
the MEGNO signature $|\Ym-2|<0.05$. About 15,700 IC lead to disruption of the
resulting configuration during the global  period of time less than
70,000~years. The results are illustrated in Figure~\ref{fig:fig7}, which
shows the fitted semi-major axis $a_c$ of the outer planet versus the fitted eccentricities
of the companions. Dots represent the fits with $\Chi<1.5$ and larger, filled
circles represent $(a_{\idm{c}},e_{\idm{c}})$ of the quasi-periodic, stable
configurations. This figure clearly shows that the stable solutions are
grouped in a few distinct  islands, the most remarkable of which is 
 the one
corresponding to the smallest $\Chi \simeq 1.473$
(about $a_{\idm{c}}\simeq 3.43$~AU). It will be shown in the
next section  that this island is related to 7:2 MMR. For all of the stable
fits, the eccentricity of the outer planet is  moderately small compared to
the previous estimates. The figure  shows also the scale of the uncertainty
of parameters form the Newtonian fits:  for the statistically similar fits,
$\Chi < 1.5$, the value of $a_{\idm{c}}$  is spread over $\simeq 2.5~$~AU,
and $e_{\idm{c}}$ over 0.6.  At the same time, $e_{\idm{b}}$ remains well
bounded.  A sharp limit of possible solutions for about $a_{\idm{c}}>2.5$~AU
is clearly visible.   All the stable fits have $1.47 < \Chi < 1.49$, rms
$\simeq 4.4~\ms$ and very similar parameters of the inner companion:
$e_{\idm{b}} \simeq 0.283$, $\omega_{\idm{b}} \simeq 134^{\circ}$,
$M_{\idm{b}} \simeq 344^{\circ}$. 

\section{Genetic fitting with the MEGNO penalty}
Although the best fit solutions obtained with the help of the previous
analysis are unstable, there is still no reason to claim that in their
neighborhood a stable fit related to a resonance is not possible. We have
already analyzed such an instance \citep{Gozdziewski2001b}. Following the
approach advocated in that paper and in a series of recent works
\citep[e.g.,][]{Gozdziewski2003c,Gozdziewski2003d},  we can
search for a stable IC by calculating and analyzing the MEGNO stability maps.
Moreover, in such a case we have to control how the tested initial condition
"preserves"  the $\Chi$ function and the shape of the RV curve. Without such
control, it is easy to  find a stable initial condition (see the analysis in
the section~2) that can lead to a very poor model of the RV observations.
Thus, for a desired (stable) solution, we have to calculate its $\Chi$ and 
decide whether it is acceptable.  In this way, we explicitly follow the
argumentation given in the introduction, i.e., the system dynamics serves as
an additional observable characterizing the planetary system. 

In practice such a procedure is very laborious so we have found a way to
simplify the search process. For this purpose, we employed the GA
minimization.  Since the GAs belong to the gradient-free optimization
methods, to find the best fit we only have to define the $\Chi$ function. To
find the  best fits that are simultaneously stable, we proceed as follows.
The ``fitness'' function $f$, required by the GA, which is equivalent to 
$1/\Chi$, is modified according to  the formula
$
  f = 1/p[\Chi,\Ym],
$
where $p$ is a ``penalty'' function. If, for the tested IC, the value of
$\Chi$ is less than a prescribed limit $\Chim$,  then the MEGNO signature
$\Ym$ is also computed for this fit. Then, if the tested fit  is related to a
regular system then $|\Ym-2| < \epsilon$  where $\epsilon>0$ is a small
value, and  $p \equiv \Chi$, i.e., $f$ is unchanged.  For a chaotic solution,
$|\Ym-2|$ grows linearly, and $f$ should diminish substantially.   The choice
of $p$ is not quite obvious. In our calculations, we tested two forms of the
penalty function:
$
 p = \Chi + \alpha |\Ym-2|,
$
and
$
 p = \Chi (1 + \alpha |\Ym-2|),
$
where $\alpha>0$. Both give similar convergence and solutions. Otherwise, if
$\Chi>\Chim$, then $f$ is left unchanged and the MEGNO test is skipped.  This
step is very relevant for the numerical  efficiency as the calculations of
MEGNO are CPU-expensive. The $f$ code depends on three control parameters:
$\Chim$, $\epsilon$, and $\alpha$. To gain the desired numerical efficiency,
we used the limit  $\Ym_{\idm{max}} \simeq 5$ for MEGNO, a relatively short
integration time (about $10^3$ periods of the outer companion), $\Chim=1.6$, 
$\epsilon \simeq 0.1$ and $\alpha =1$. This integration time is sometimes too
short to get a clear MEGNO convergence, but the code quickly eliminates the
strongly chaotic systems. This also weakens the requirement of strictly
regular, quasiperiodic configurations, and  the search does not exclude
systems evolving on a border of stability. Obviously, in this approach,  the
RV model is driven by the full model of the $N$-body dynamics.  Because the
search is driven by the GA, the MEGNO-penalty algorithm has also a global
character.

Using the described method in a number of repeated runs of the code,  the
algorithm converged repeatedly to a few distinct solutions, which are called
GM fits from hereafter.   Their examples, labeled GM1-GM6, are given in
Table~{\ref{tab:tab3}.  At first we limited the search region to 
$a_{\idm{c}} < 3.6$~AU. In this case one could expect solutions
corresponding to  the lower order resonances, including the 2:1~MMR. Indeed, we
found two such different fits: the one corresponding to the 3:1~MMR (GM1,
$a_{\idm{c}}$ about 3.1~AU) and the second  one related to the 7:2~MMR (GM2,
$a_{\idm{c}}$ about 3.44~AU). The GM2 fit appears to be the best stable fit
found in the entire search, having $\Chi \simeq 1.47$ and rms $\simeq
4.35~\ms$.  At the same time, we did not find any stable solution that would
correspond to the 2:1~MMR.   In the region $3.6~\mbox{AU} < a_{\idm{c}} <
5.2~\mbox{AU}$, we found a number of equally good fits with $\Chi \simeq
1.48$ and rms $\simeq 4.4~\ms$.  The dynamical environment of the GM fits  is
shown as MEGNO scans in the $(a_{\idm{c}},e_{\idm{c}})$-plane
(Figure~\ref{fig:fig8}). Note, that the stable solutions have moderately
small $e_{\idm{c}} \simeq 0.3$-$0.5$  compared to the previous estimates.  
In some cases, due to the short time of the MEGNO integrations, the algorithm
finds solutions residing close to the border of stability. Good examples of
these are GM1 and GM3. Nevertheless, these fits are very close to the areas
of stability. Finally,  for a reference, the synthetic RV curve for the best
GM2 fit is shown in the right panel of Figure~\ref{fig:fig6}.  The synthetic
RV curves of the other GM fits  have very similar shapes, thus are not shown
there. Let us also note that the MMRs  structures apparent in the MEGNO-maps
are interestingly similar to those obtained for the outer Solar system with
the Frequency Analysis scheme \citep{Robutel2001}. 

The results of the global GM search are in accord with the results of the
quasi-global $N$-body fit described in section~3.  Figure~\ref{fig:fig7}
shows the localization of the GM fits in the
$(a_{\idm{c}},e_{\idm{c}})$-plane of the best Newtonian fits. Remarkably, the
GM search leads to the same best solution related to the 7:2~MMR. The other
fits coincide very well too. This is a very relevant conclusion as both
kinds  of the fits are obtained through completely independent algorithms. It
confirms the reliability of the GM fits and, actually, the NL2 fits.   

Finally, we computed the evolution of the orbital elements, for all the GM
fits, given in Table~\ref{tab:tab3}. The results are illustrated in
Figures~\ref{fig:fig9} and~\ref{fig:fig10} and show the complexity of the
possible dynamical  behaviours of the \pstar system that are consistent with
the RV observations.  The fits GM1--GM4 correspond to MMRs: 3:1, 7:2, 9:2 and
14:3, respectively. As we already mentioned, the GM3 fit is unstable although
it lies close to a stable region. In this case, a curious SAR with
$\theta_1=\varpi_{\idm{b}} +\varpi_{\idm{c}}$ librating about $90^{\circ}$
appears. For the other ICs, the SARs with different libration centers are
also present. In the GM1 fit, corresponding to the 3:1 MMR, $\theta$ librates
about $240^{\circ}$. The rest of the GM fits have $\theta$ librating about
$180^{\deg}$, with the semi-amplitudes as small as a few degrees (see, e.g.,
Figures~\ref{fig:fig9}d and \ref{fig:fig10}b). Note also the different
evolution of the eccentricities of the companions and the stabilizing
influence of the MMRs. 

%
%
\section{Previous dynamical studies of the \pstar system}
%
%
The \pstar system has attracted the attention of many researchers, most
likely due to its apparent proximity to the 2:1 MMR. We have found very
interesting to compare the results of previous studies with the current and
still very fragmentary knowledge of its dynamics.

Shortly after the reprint by \cite{Jones2002} had appeared,
\cite{Kiseleva2002} used the preliminary fit J2 to analyze its stability. Using
the MEGNO technique, they found stable configurations located not very far 
in the parameter space from this initial condition and concluded that high
eccentricity of the outermost  planet $e_\idm{c} > 0.7$ is an important
stabilizing  factor in the \pstar system. A similar conclusion was quoted by
these authors in \citep{Kiseleva2002a}.  In the light of our study, these
results are not quite clear for us. In our MEGNO  maps, in the same
($e_{\idm{b}},e_{\idm{c}}$)-plane, computed for the J2 initial condition with
the resolution $100 \times 100$ points, there are not any stable points for 
$e_{\idm{c}}>0.1$.  It is illustrated in the left panel of 
Figure~{\ref{fig:fig11}}. All configurations with high $e_{\idm{c}}$ are
self-disrupted.

In fact, in their Table~1, \cite{Kiseleva2002} quote a different time of the
periastron passage of the inner planet than the one given in \cite{Jones2002}
and  \cite{Jones2003}. Indeed, for IC changed this way,  (see the fit K2 in
Table~\ref{tab:tab1}), we also obtained a very extended, ridge-like zone of
stability for  extremely large  $e_{\idm{c}}$ (see the middle panel in
Figure~\ref{fig:fig11}). Moreover, the initial configuration resulting from
that modified IC  produces the RV signal in anti-phase with the original RV
curve---compare the synthetic RV curves obtained for the original and the
modified IC, respectively (shown in Figure~\ref{fig:fig13}).  Actually, it
seems that all the double-Keplerian RV curves shown in Figure~\ref{fig:fig13}
do not describe even approximately the measurements.  We did not succeed in
reproducing the double-Keplerian RV curves, which have been defined by the
literal values of the elements $K,n,e,\omega,T_{\idm{p}}$ of the J2 and J2a
fits (see Table~\ref{tab:tab1}), and calculated with the classical formulae
\citep[see][]{Smart1949,Lee2003}, as they have been shown in \cite{Jones2003}
in their Figure~3 (see our Figure~\ref{fig:fig13}). Only the one-Keplerian
signal of the inner planet (given by the J2a fit) comes reasonably close to
the RV data points. Possibly, in the original Figure~3 there is shown  the
one-Kepler+trend signal. Anyway, this shows that the literal elements
of the outer planet in
the J2 and J2a fits cannot be treated as representative for the Doppler
signal of the \pstar.

Further, while trying to explain the differences between the results of
\cite{Kiseleva2002} and this paper, we noticed some stable points in their
Figure~3d, located about  $e_{\idm{b}} \simeq 0.2$ which are absent in our
version of this scan (see the middle panel of Figure~\ref{fig:fig11}). We see
at least two possible explanations of this difference. The quoted authors did
not specify the initial epoch of their integrations, so the different epochs
can give different osculating elements and finally the MEGNO signatures. This
can be also related to a ``pathological'' MEGNO  convergence that is
characteristic for collisional  orbits. The Lyapunov exponent-based criterion
for chaos is misleading if bodies are ejected from the system
\cite[]{Lissauer1999}, because after the ejection the distance between the
phase trajectories no longer grows exponentially. The ejected planet can stay
on a distant orbit and apparently  the system is bounded and regular. In such
cases MEGNO can converge very close to 2, finally giving a correct
identification of a stable configuration.  As an example we show the results
of the integration of the \pstar system when the J2 is changed: $e_{\idm{b}}$
from 0.31 to 0.25 and $e_{\idm{c}}$ from 0.8 to 0.83. The resulting MEGNO
behaviour is depicted in Fig.~{\ref{fig:fig12}}. The indicator converges to
2.05 in a way characteristic for a regular system. Nevertheless, from the
qualitative point of view, the resulting system is very different from the
starting configuration.  Thus, following \cite{Lissauer1999}, we treat the
ejections as strongly chaotic phenomena, in spite of the apparent regularity
of the final orbital states. Not eliminating this effect can lead to false
patterns in the MEGNO stability maps that can be even recognized as narrow
resonance zones. We noticed and fixed this problem with the MEGNO convergence
in \citep{Gozdziewski2001b}.

In a recent paper, \cite{Bois2003} deal with a  global analysis of the 2:1
MMR, assumed to be present in the \pstar system.  These authors report the
initial condition, which is equivalent to the J2 fit but modified in the same
way as the K2 fit, and additionally, in this IC, $a_{\idm{b}}$ has been
changed from 2.3~AU to 2.381~AU. Then, using the MEGNO indicator,  they
analyze the dynamics of resulting planetary configurations and compare them
with a similar case of the Gliese~876 2:1 MMR. In the neighborhood of the
analyzed IC, there is an extended stable zone of this resonance. However,
also in this case the system dynamics that has been studied, does not
reproduce the RV measurements of the \pstar.
%
%
\section{Conclusions}
%
%
In this paper we  attempt to verify the hypothesis stated by \cite{Jones2003}
that the linear trend apparent in the RV observations of \pstar is a Doppler
signal of the second planetary companion. The problem we have to deal with
while trying to estimate the orbital elements of the hypothetical planet, is
a very short timespan of the data compared to the probable orbital period of
the outer planet. Our analysis incorporates the already well recognized fact
that the commonly used Keplerian model of the RV data can falsify the
dynamics of the studied system. This statement has been stated by
\cite{Laughlin2001} who pioneered the Newtonian (or ``self-consistent'',
i.e., incorporating the mutual interaction between companions) RV fits and
have applied them to strongly interacting  planets around Gliese~876. The
\pstar system is yet another  example that neglecting the full dynamics in
the fitting process of the orbital elements to the RV observations leads to
artifacts when the RV data permit elongated orbits  and relatively small
orbital distances between the planets.  For the \pstar system, the formally
best Keplerian fits describe configurations that lead to a collision of the
companions during a few weeks. But such configurations are not likely as we
would be observing the system just after or just before the collision event.
In this sense the dynamics become an important observable that has to be
taken into account with the same priority as the RV observations. To give
this idea a numerical implementation, we propose a method that merges the
genetic optimization algorithm with the MEGNO analysis, i.e., the examination
of the orbital stability by a fast indicator. The efficiency of the MEGNO
algorithm makes it possible to automate the fitting process and to gain an
acceptable overall numerical efficiency of the algorithm.  With such a
MEGNO-penalty fitting approach,  we find a number of stable orbital fits
having $\Chi \simeq 1.48$ and an rms $\simeq 4.4~\ms$. The search have been
confined to the limit of of the Jupiter-like periods of the hypothetical
second planet. These configurations correspond to the low-order MMRs 3:1, 7:2
and to the neighborhoods of 14:3, 4:1, and 5:1 MMRs.  This result is in
excellent accord with the quasi-global, gradient search of the best Newtonian
fits. It demonstrates that our algorithm reliably finds the desired,
stable initial conditions that produce a synthetic Doppler signal consistent
with the observations.

The current set of the RV data does not constrain the orbital elements of
the  outer planet. The preliminary Keplerian fits lead to a completely false
dynamical representation of the \pstar system. Considerably more significant
constraints  on the possible orbital configurations and parameters of this
system are provided by the dynamics. Our paper proves that the future
analysis of the updated RV observations has to be driven by the full,
$N$-body model of the RV observations.

Forecasting the real state of the \pstar system is very difficult now. The RV
data permit a continuum of statistically equivalent orbital fits. However,
the extensive dynamical analysis of these solutions makes it possible to give
some overall characteristics of the planetary system even at this stage.   If
the outer planet revolves close to the inner companion, it is unlikely to
maintain the system stability for  large $e_{\idm{c}}$ without a stable
orbital resonance.
For a distance up to about $3$~AU, the system  can be locked in the low-order
MMRs 3:1 or 7:2. For larger values of the semi major axis of the outer
companion, up to the Jupiter-like $5.2$~AU, the system can be found in an
extended zone confined to other low order MMRs, like 4:1, 5:1 or to their
neighborhoods,  accompanied by the SAR with semi-major axes antialigned in
the exact resonance. However this type of the secular resonance does not
exhaust other possibilities. Although it is speculative, the dynamical
structure of this region reminds that of the HD~12661 system. Possibly, the
large libration island of the SAR with the apsidal lines anti-aligned in the
exact resonance can be explained by the secular theory of \cite{Lee2003}.
Such a test is not straightforward to carry out because every data point in
the scan represents a different initial condition and the scan data pass into
the neighborhoods of the low-order MMRs. The dynamical similarity of both
systems flows also from their RV signals and the masses of their host stars:
if one would restrict the period of observations of the HD~12661 system to
the range between JD=24511200 and JD=24511800 
\citep[see the RV data in][]{Fischer2003}, then we
would see the same kind of a linear trend visible in the RV data of the
\pstar. Also the MEGNO maps of the neighborhoods of the best stable GM fits 
and for the best fits of the HD~12661 system \cite[]{
Gozdziewski2003c,Gozdziewski2003d} reveal
many similarities. 

Because our study has mostly qualitative character,  we omitted some effects
in the fitting process that should be taken into account, when a more
extended set of the RV data will be accessible. Specifically, we did no
discuss the formal and systematic errors (e.g., from  the stellar RV jitter) 
of the best fit parameters. This in principle could be done as in our  recent
paper devoted to the HD~12661 system \cite[]{Gozdziewski2003d}.  We only
considered coplanar and edge-on  configurations. Incorporating the mutual 
interaction between planets would remove the geometrical degeneracy which
does not  permit to estimate the relative inclination of the orbits and the
inclinations  of the orbital planes \cite[]{Laughlin2001,Rivera2001}.
Currently, the small  number of measurements is an obstacle to  perform such
fully self-consistent fits.  Hopefully, the   data set will steadily grow and
soon will be sufficiently large  to verify of our approach and predictions.

\section{Acknowledgments}
Calculations in this paper have been performed on the HYDRA
computer-cluster,  supported by the Polish Committee for Scientific Research,
Grants No.~5P03D~006~20 and No.~2P03D~001~22. This work is supported by the
Polish Committee for Scientific Research, Grant No.~2P03D~001~22. K.G. wants
to acknowledge the support by the  N. Copernicus University, Grant No~362-A.
M.~K. is a Michelson Postdoctoral  Fellow.
%
%
%
\section*{Appendix: A model of the radial velocity observations}
%
%

Below we describe the Keplerian model employed in this paper.
Our model differs in some details from the commonly used version
which comes from the classical RV studies of stellar binaries
\citep[see e.g.][]{Smart1949}.

We start with the basic definitions of the Keplerian motions. The radius vector
$\bR(t)$ and the velocity$\bV(t)$  of a body moving in an elliptic orbit are 
given by %
\begin{gather}
\label{e:RtKepler}
\bR(t) =  a\left[(\cos E(t) - e)\,\bP  + \sqrt{1 - e^2}\sin E(t)\,\bQ \right], \\
\label{e:VtKepler}
   \bV(t) = a n \left[ -\frac{\sin E(t)}{1-e \cos E(t)}\bP  +
 \frac{\sqrt{1 - e^2}\cos E(t)}{1-e \cos E(t)}\bQ\right], 
\end{gather}
where 
%
$
\bP = \bl\,\cos\omega + \bm\,\sin\omega,
$
$
  \bQ = -\bl\,\sin\omega + {\bf m}\cos\omega,
$
and
%
\[
  \bl = \begin{bmatrix}\cos\Omega \\
                     \sin\Omega \\
                       0
\end{bmatrix}, \qquad
  \bm = \begin{bmatrix}-\cos i\sin\Omega \\
                     \phantom{-} \cos i\cos\Omega \\
                     \phantom{- \cos \Omega} \sin i 
         \end{bmatrix}.
\]
The eccentric anomaly $E=E(t)$ is an implicit function of time 
through the Kepler equation
$
  E - e\sin E = M,
$
where $M$ is the mean anomaly
$
M = n(t - T_{\mathrm{p}}),
$,
$
n = 2\pi/P,
$
and $P$ is the orbital period of the body. The remaining parameters
$a$, $e$, $\omega$, $\Omega$, $T_p$ are the standard Keplerian
elements --- semi-major axis, eccentricity, longitude of pericenter,
longitude of ascending node and time of pericenter. The mean motion
$n$ and the semi-major axis $a$ are connected by the relation
$ 
   n^2 a^3 = \mu,
$
where $\mu$ is the gravitational parameter. The explicit form of $\mu$
and the meaning of $\bR(t)$, $\bV(t)$ depend on which Kepler problem
we consider (relative orbit in the two body problem, barycentric orbit,
etc).

Now, let us assume that the body is a planet of mass $m$ revolving around a
star of mass $m_\star$, and let $\bR(t)$, $\bV(t)$ and $\bR^\star(t)$,
$\bV^\star(t)$ denote the \emph{barycentric} radius vector and the
velocity of respectively the planet and the star. By $\brho(t) :=
\bR(t)- \bR^\star(t)$ and $\bupsilon(t) := \bV(t) - \bV^\star(t)$ 
we denote the \emph{relative} radius vector and 
velocity of the planet. From the definition of the center of mass we have
\begin{gather}
\label{eq:Rtstar}
 \bR^\star(t) = -\frac{m}{m_\star}\bR(t)= -\frac{m}{m_\star+m}\brho(t),\\
\label{eq:Vtstar}
 \bV^\star(t) = -\frac{m}{m_\star}\bV(t)= -\frac{m}{m_\star+m}\bupsilon(t),
\end{gather}
We choose the barycentric reference frame in such a way that its third
axis has the direction of the vector from the observer to the mass center
of the system. Hence the radial velocity of the star is the third
component of its barycentric velocity, 
%
$v^\star_{\mathrm{r}} (t) := V^\star_3(t)$.
%
Now, we can express $v^\star_{\mathrm{r}} (t)$ in terms of Keplerian elements of the planet:

\begin{equation}
\label{eq:vrtexp}
v^\star_{\mathrm{r}} (t) = -L \left[\frac{\sqrt{1 - e^2}\cos E(t)\cos \omega -\sin E(t) \sin \omega}{1-e \cos E(t)}\right], 
\end{equation}
where the explicit form of $L$ as well as all Keplerian elements
depend on our choice of the planetary orbit
\footnote{Note the ``minus'' sign before $L$. The classic expression of
the RV signal is given in terms of the true anomaly $\nu$.
For details see \cite{Smart1949} or a recent paper by \cite{Lee2003}
}. 

In the case of the
barycentric orbit
%
$
L = \sigma  a n \sin i, 
$ 
where $\sigma=m/m_\star$ and 
$
\mu= G m_\star^3/(m_\star+m)^2.
$
For the relative orbit the form of $L$ is the same, with
$\sigma=m/(m_\star+m)$ and
$
\mu= G( m_\star + m).
$
%
%
It should be noted here that not only semi-major axes calculated for the
barycentric and the relative orbit of the planet are different but also the
eccentricities may be different. 

When performing the least-squares fit to the radial velocity data both
parameterizations give the same results (of course, after an appropriate
transformation). It is not so clear when we have more than
one planet. Typically, the radial velocity of the star is modeled in
barycentric coordinates as a sum of terms \eqref{eq:vrtexp}  calculated for
each planet (but the semi-major axis of planet's orbits and  the lower limits
on their masses are then often given for the relative orbits which is a
strange mismatch for systems containing more than one planet).
However in barycentric coordinates, even if we assume that  the planets do
not interact directly, their orbits are not Keplerian.  In such a case, the
companions still interact indirectly via the host star. Only in the Jacobi
coordinates the direct and indirect effects of the mutual  interactions
between planets are much smaller than the leading terms of  the star-planet
interactions. 
 Note that this is a well known fact---for instance,
the already classic  symplectic integrators  by \cite{Wisdom1991} heavily
rely on the use of Jacobi coordinates.
Recently, this has been analyzed in the context of the RV data modeling
in  by \cite{Lee2003}. They point out that the double-Keplerian
orbital fits have to be expressed in Jacobi coordinates and not in  the
barycentric coordinates.  Further, using the ``classic''
approach, we are not allowed to calculate the masses of the companions  by
using the same formulas as for the barycentric motion of the two bodies.
There is not any well defined gravitational parameter if the RV fits of a
multi-planetary system are represented directly in terms of the barycentric
Keplerian elements.  Amazingly,  these actually quite obvious conclusions are
relevant to tens of already published papers on extrasolar planets discovered
through precise RVs! It is worth to note  that in recent years, in the
extrasolar planet field it has been  properly approached in the case of
PSR~B1257+12 planetary system \citep{Konacki2000}. \cite{Lee2003} 
give a clear explanation  of the problem in the context of RV 
measurements.

For simplicity, let us assume that we have two planets with the masses $m_1$
and $m_2$. Their barycentric radius vectors an velocities we denote by
$\bR_i$, $\bV_i$, $i=1,2$. The Jacobi coordinates and velocities $\br_i$,
$\bv_i$ of the planets are defined in the following way
\begin{gather}
\label{eq:rjac}
\br_1 := \bR_1 - \bR^\star, \qquad 
\br_2 := \bR_2 - \frac{1}{m_\star + m}\left( m_\star\bR^\star + 
                                             m_1\bR_1 \right),\\
\label{eq:vjac}
\bv_1 := \bV_1 - \bV^\star, \qquad 
\bv_2 := \bV_2 - \frac{1}{m_\star + m}\left( m_\star\bV^\star +
                                             m_1\bV_1 \right).
\end{gather}
Thus, $\br_1$ is the position of $m_1$ relative to $m_\star$, and $\br_2$ is 
the position of $m_2$ with respect to the center of mass of $m_\star$ and 
$m_1$. In terms of the Jacobi coordinates we have 
\begin{equation}
\label{eq:RVstarJ}
\bR^\star = -\sigma_1 \br_1 - \sigma_2\br_2, \quad 
\bV^\star = -\sigma_1 \bv_1 - \sigma_2\bv_2,
\end{equation}
where 
%
$
\sigma_1 = m_1/(m_\star+m_1)
$, 
$\
\sigma_2 =  m_2/(m_\star+m_1 + m_2).
$
%
As we already mentioned, in the Jacobi coordinates we can consider the
three body problem as a perturbation of two the two body models:
the planet $m_1$ plus the star, and the planet $m_2$ plus a fictitious 
point of mass $m_\star + m_1$.  Thus, approximating the planetary motion 
by these two problems, we obtain that
$v^\star_{\mathrm{r}} (t) = -L_1f_1(t) -L_2f_2(t)$, where
$L_k =\sigma_k a_k n_k \sin i_k$, 
$k=1,2$,
$\mu_1= G( m_\star + m_1)$,  
$\mu_2= G( m_\star + m_1 + m_2)$,
and
\begin{equation}
\label{eq:fkt}
f_k(t) = \frac{\sqrt{1 - e_k^2}\cos E_k(t)\cos \omega_k -
\sin E_k(t) \sin \omega_k}{1-e_k \cos E_k(t)}.
\end{equation}

\bibliographystyle{apj}
\bibliography{ms.bib}
\newpage
%
%
\begin{table}
\caption{
Orbital parameters of the \pstar system, adopted from data published in
\cite{Jones2002} (J2), \cite{Jones2003} (J2a), 
and used in \cite{Kiseleva2002}(K2).  
The mass of the central star is equal to
$1.08~\mbox{M}_{\sun}$.
}
\smallskip
\begin{tabular}{lcccccc}
\hline
 & \multicolumn{2}{c}{J2} 
 & \multicolumn{2}{c}{J2a} 
 & \multicolumn{2}{c}{K2} \\
\hline
Orbital parameter / planet  & b &  c  
&  b &  
                                &  b &  c  \\
\hline
$\mbox{m}_{\idm{pl}} \sin i$ [M$_{\idm{J}}$] \dotfill &  1.7 & (1.0) & 1.7 &
($>$1.5) &  1.7  &  1.0 \\
$a$ [AU] \dotfill & 1.5 & (2.3)  & 1.5 & ($>$2.5) &  1.5  & 2.3   \\
$P$ [d] \dotfill & 638 & (1300) & 637 & (1500) &   638 & 1300 \\
$e$ \dotfill     &  0.31 & (0.8) & 0.31 & (0.8) &  0.31 & 0.8 \\
$\omega$ [deg]\dotfill &  320 & (99) & 320 & (99) & 320 & 99 \\
$T_{\idm{p}}$ [HJD-2400000]  \dotfill&  50958  & (51613)  & 50959 &  (51613) 
  & 50698 & 51613 \\
K [$\ms$]      & 40 & (34.2) & 40 & (34) \\
rms [$\ms$]      & 5.42 & (5) & 5.28 & (5) \\
\hline
\end{tabular}
\label{tab:tab1}
\end{table}

\begin{table}
\caption{
The orbital parameters of the \pstar system derived from the double-Keplerian
Levenberg-Marquardt fit (JK2), the genetic fit (GA2), and the
Levenberg-Marquardt self-consistent Newtonian fit (NL2). The JK2 and GA2
parameters are  related to Jacobi coordinates; the NL2 fit is given in the
astrocentric Keplerian elements at the epoch of the first observation ($t_0
\equiv$ JD2450915.2911). The mass of the parent star is equal to
$1.08~\mbox{M}_{\sun}$. See the Appendix for an explanation of the symbol
$L$.
}
\smallskip
\begin{tabular}{lcccccc}
\hline
 & \multicolumn{2}{c}{JK2} & \multicolumn{2}{c}{GA2} & \multicolumn{2}{c}{NL2}\\
\hline
Orbital parameter /planet  &   b  &  c 
&  b  &  c  &  b  &  c  \\
\hline
$\mbox{m}_{\idm{pl}} \sin i$ [M$_{\idm{J}}$] \dotfill 
               &  1.68 & 2.09  &  1.69   &  1.44 & 1.68 & 2.18 \\
$L$ [$\ms$]  \dotfill     & 38.2 & 34.7 & 38.3 & 23.5 &  & \\
$a$ [AU] \dotfill & 1.449 & 2.708  & 1.460 & 2.813 & 1.437 & 2.724 \\
$P$ [d] \dotfill & 612.4 & 1563.6 & 619.43 & 1655.5 &  & \\
$e$ \dotfill     & 0.348 & 0.864 & 0.326 & 0.700 & 0.340 & 0.879 \\
$\omega$ [deg]\dotfill & 141.0  & 311.4 & 140.1  & 293.4 & 141.7 & 314.4 \\
$T_{\idm{p}}$ [JD-2450000]  \dotfill& 393.4  &  940.5  &  2231.2  & 2601.1 \\
$M(t_0)$ [deg] \dotfill & 306.8 &  354.2 & 315.2 & 353.4 & 307.7 & 354.7 \\
$V_0$ [$\ms$] & \multicolumn{2}{c}{-10.9} & \multicolumn{2}{c}{-6.6} & 
   \multicolumn{2}{c}{-11.0}       \\
$\Chi$  & \multicolumn{2}{c}{1.38} & \multicolumn{2}{c}{1.41} & 
   \multicolumn{2}{c}{1.40}                             \\
rms [$\ms$] & \multicolumn{2}{c}{3.9} & \multicolumn{2}{c}{4.0} & 
    \multicolumn{2}{c}{4.0}        \\
\hline
\end{tabular}
\label{tab:tab2}
\end{table}

\begin{table}
\caption{
Newtonian, genetic fits with the MEGNO penalty test. Osculating, astrocentric
Keplerian elements are given for the the epoch of the first observation
(JD=2450915.2911).
}
\smallskip
\begin{tabular}{ccccccccccl}
\hline
Fit&Planet&$V_0$&$\Chi$&rms& $m_{\idm{pl}}\sin i$ & $a$ & $e$ & $\omega$ & $M$ &
Note \\
   &   &  [$\ms$] & & [$\ms$] & [$m_{\idm{J}}$] & [AU] & & [deg] & [deg] & \\
\hline
\hline
GM1 & a & -7.2 & 1.49 & 4.40 & 1.69 & 1.493 & 0.287 & 133.0 & 344.1  &  \\
    & b &      &      &      & 1.26 & 3.117 & 0.457 & 290 & 33.6 & SAR ($240^{\circ}$) \\
stable& b &      &      &      & 1.26 & 3.100 & 0.457 & 290 & 33.6 & 3:1 MMR \\
\hline
GM2 & a & -5.1 & 1.47 & 4.35 & 1.72 & 1.493 & 0.284 & 133.2& 344.2 & 7:2 MMR \\
(best)    & b &      &      &      & 1.32 & 3.444 & 0.377 & 302.1& 26.0  & SAR ($180^{\circ}$) \\
\hline
GM3 & a & -7.4 & 1.48 & 4.40 & 1.70 & 1.493 & 0.284 & 133.5 & 343.4 & $\simeq$ 9:2 MMR \\
    & b &      &      &      & 1.71 & 4.048 & 0.304 & 307.1 & 53.4 &  \\
\hline
GM4 & a & -4.1 & 1.48 & 4.40 & 1.70 & 1.491 & 0.283 & 135.9 & 340.0 & 14:3 MMR\\
    & b &      &      &      & 1.52 & 4.178 & 0.250 & 325.1 & 24.4  & SAR ($180^{\circ}$)\\
\hline
GM4a & a & -4.3 & 1.49 & 4.35 & 1.73 & 1.492 & 0.285 & 137.9 & 338.7 & \\
    & b &      &      &      & 1.44 & 4.144 & 0.210 & 332.3 & 17.5  & SAR ($180^{\circ}$)\\
\hline
GM5 & a & -3.9 & 1.48 & 4.36 & 1.74 & 1.494 & 0.289 & 134.8 & 343.4 &  \\
    & b &      &      &      & 1.52 & 4.305 & 0.236 & 329.1 & 25.0  & SAR ($180^{\circ}$)\\
\hline
GM6 & a & -3.7 & 1.48 & 4.33 & 1.73 & 1.493 & 0.289 & 135.4 & 342.1 & \\
    & b &      &      &      & 1.67 & 4.773 & 0.220 & 347.8 & 16.3 & SAR ($180^{\circ}$) \\
\hline
\end{tabular}
\label{tab:tab3}
\end{table}

\newpage

\begin{figure*}[th]
\centering
          \includegraphics[]{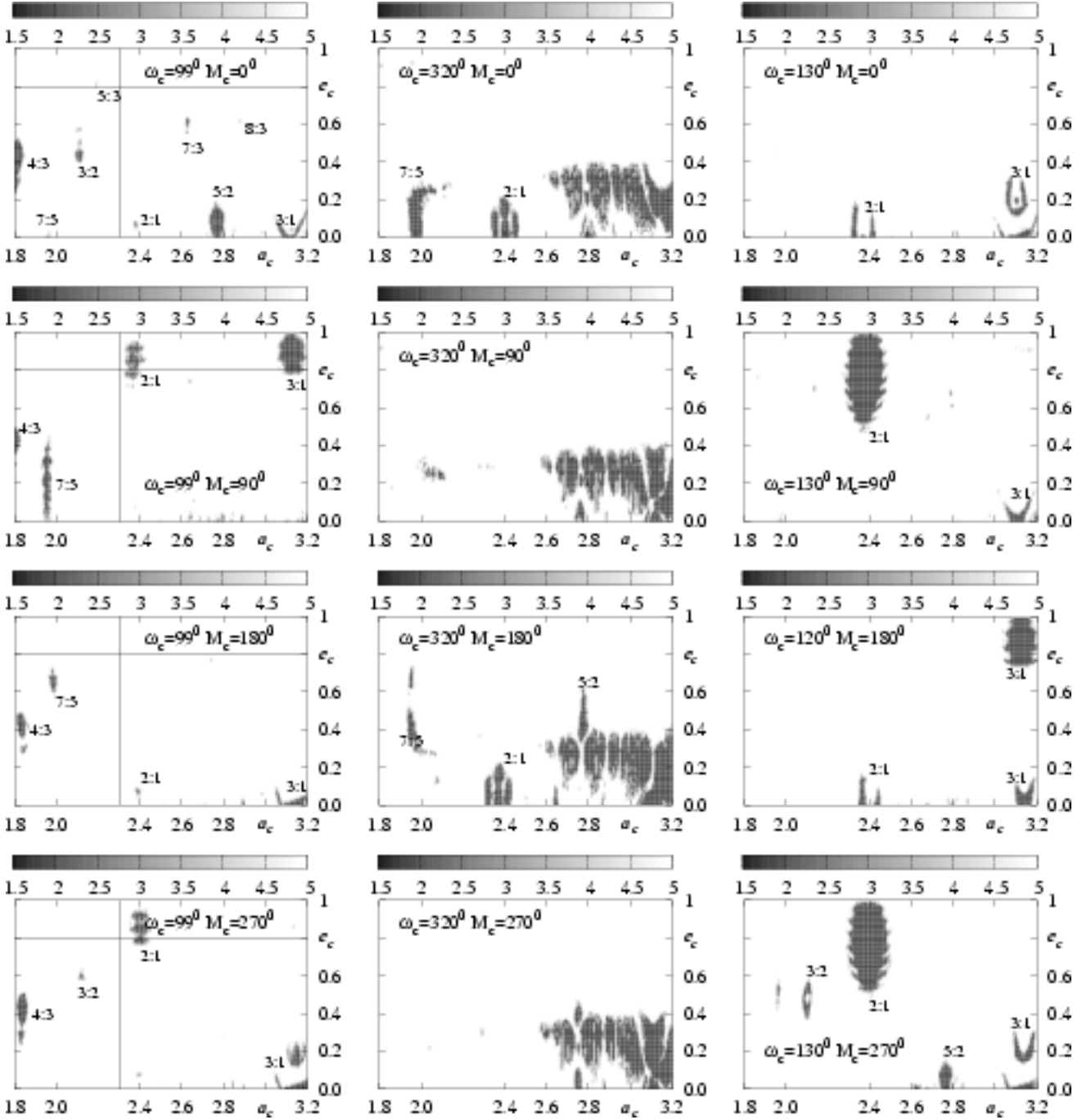}
\caption{
Stability MEGNO-maps in the ($a_{\idm{c}},e_{\idm{c}})$-plane
for different initial orbital phases of the outer planet.
The orbital elements are related to the J2 fit,  see Table~\ref{tab:tab1}. 
The left
column corresponds to the nominal $\omega_{\idm{c}}$ of the outer planet.
Some of the stable areas (shaded)
corresponding to the mean motion resonances are labeled. The resolution of the
plots is $240 \times 100$ points.
}
\label{fig:fig1}
\end{figure*}

\begin{figure*}[th]
\centering
\hbox{    
  \includegraphics[]{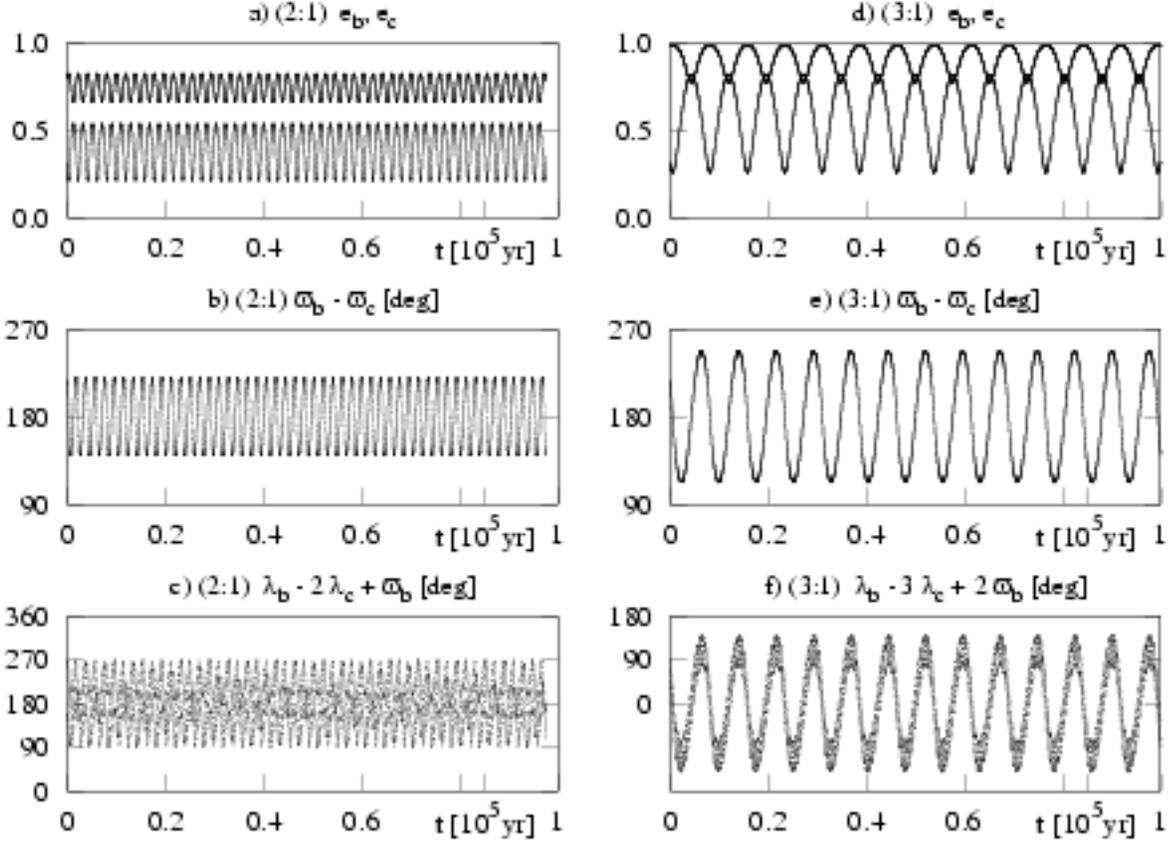}
        }
\caption{
Evolution of the Keplerian, astrocentric elements in the 2:1 MMR (the left
column, $a_{\idm{c}}=2.36$~AU, $e_{\idm{c}}=0.792$, $M_{\idm{c}}=90^{\circ}$)
and in the 3:1 MMR (the right column, $a_{\idm{c}}=3.1416667$AU,
$e_{\idm{c}}=0.9801$, $M_{\idm{c}}=90^{\circ}$).  Note that these parameters 
are modified orbital parameters of the nominal J2 fit, interpreted here as the
osculating  Keplerian elements at the epoch of the periastron passage of the
outer planet. They  are  localized in the zones of stability shown in
Figure~\ref{fig:fig1}. 
}
\label{fig:fig2}
\end{figure*}

\begin{figure*}[th]
\centering
 \includegraphics[]{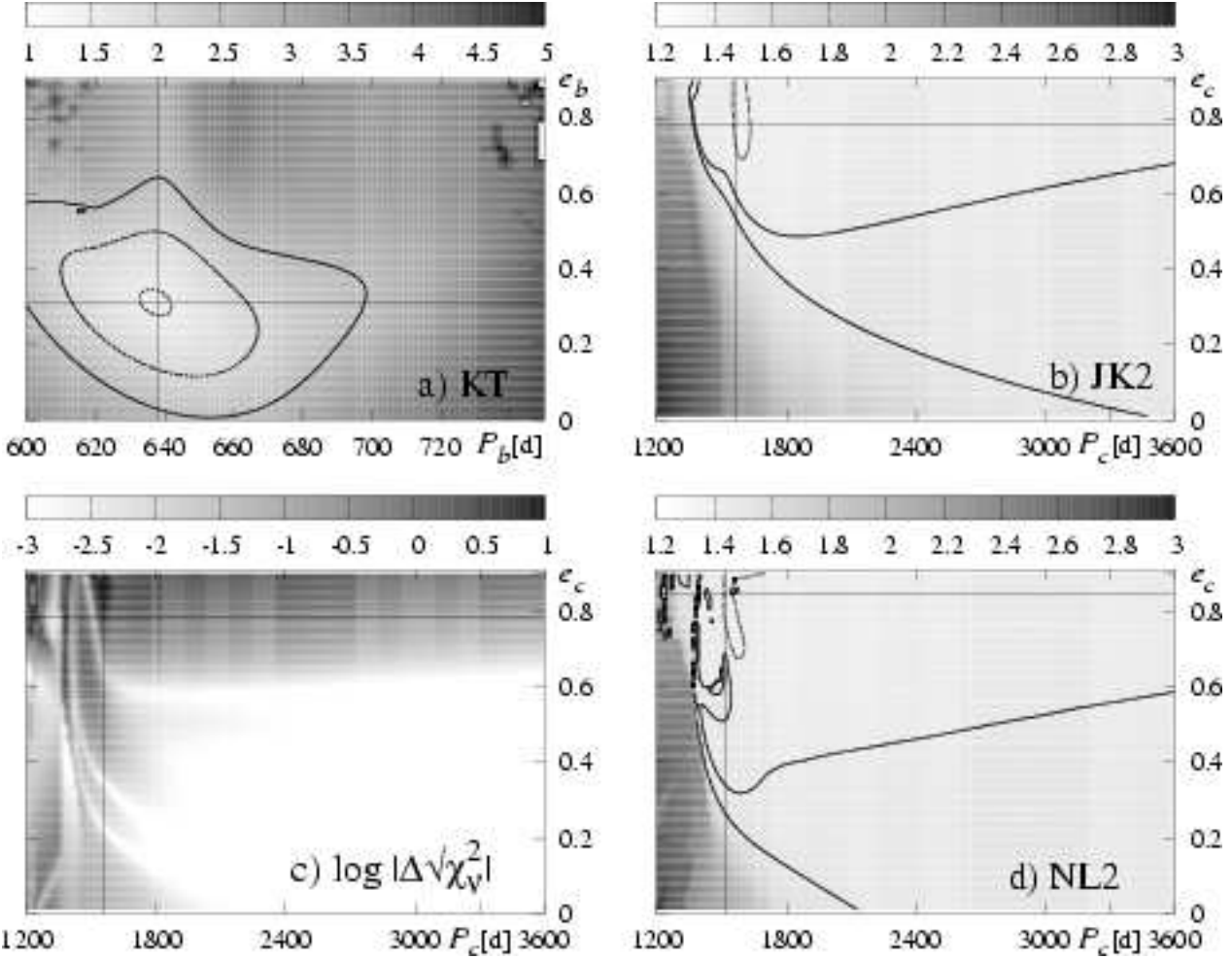} 
\caption{
Panel a) is for the function $\Chi(P_{\idm{c}},e_{\idm{c}})$ of the RV  model
which incorporates one planet in a Keplerian orbit and a linear trend. 
Contour levels are: 1.5, 2.0, 2.5. Panel b)  is for the double-Keplerian
model of the RV data and its $\Chi(P_{\idm{c}},e_{\idm{c}})$. Contour levels
are: 1.40, 1.45, 1.5. Panel c) is for the logarithm of the absolute difference
in $\Chi(P_{\idm{c}},e_{\idm{c}})$ between the double-Keplerian model and the
Newtonian model of the RV data. The best double-Keplerian solution is marked
by the the intersection of the two straight lines. Panel d) is a map of
$\Chi$ computed from the self-consistent Newtonian model of the dynamics 
when the orbital parameters are fitted by the gradient method, starting
from the best double-Keplerian solutions (see the text for more details).  
Contour levels are: 1.40, 1.45,
1.5. The resolution of the plots  is $200 \times 100$ points.
}
\label{fig:fig3}
\end{figure*}

\begin{figure*}[th]
\centering
          \includegraphics[]{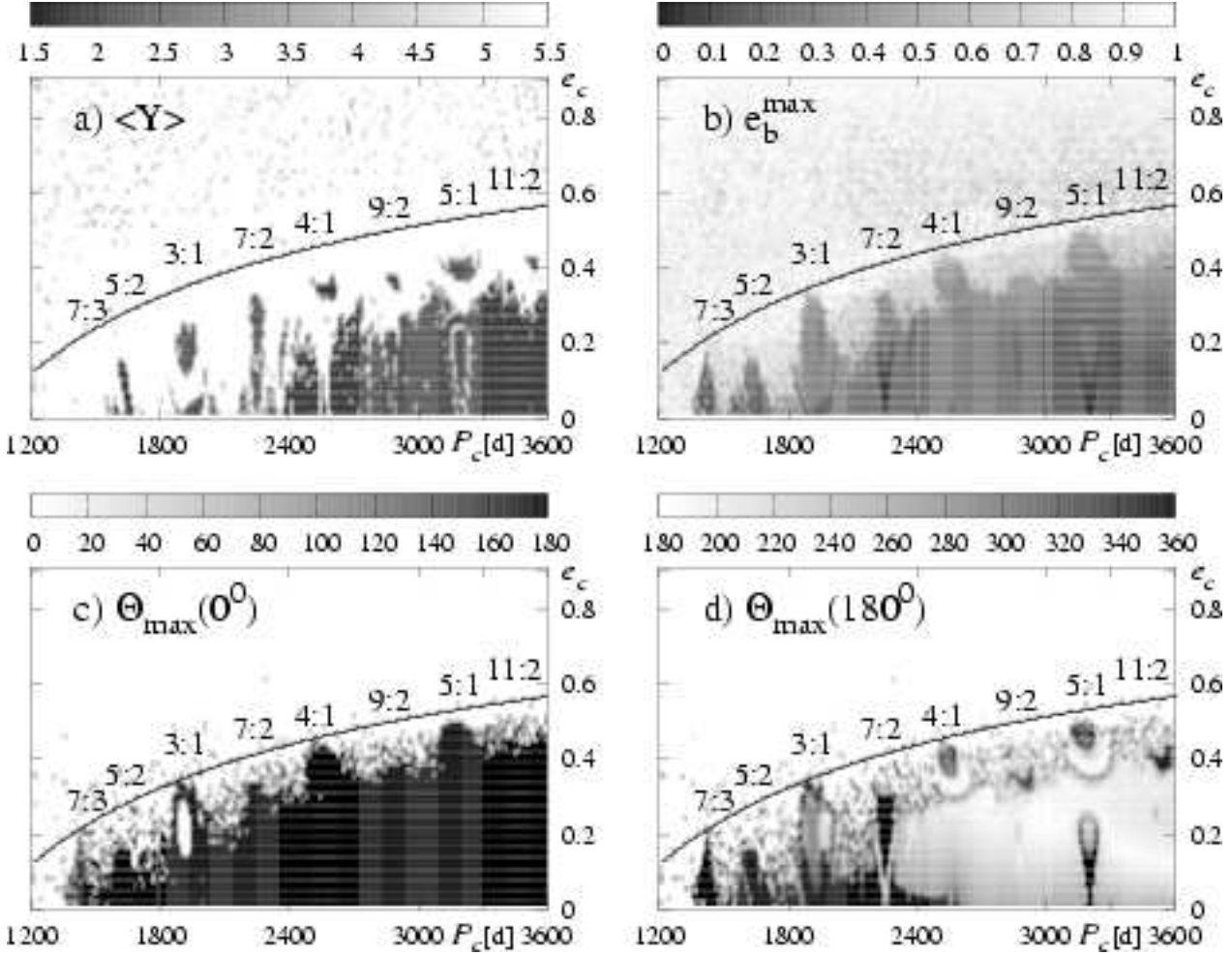}
\caption{
Dynamical scans of the  $\Chi$-map of the best double-Keplerian fits 
(Figure~\ref{fig:fig3}b). The orbital parameters of
these fits have been mapped to the osculating astrocentric elements at the
epoch of the first observation. The top left panel 
(a) is for the MEGNO, the top
right panel (b) is for the maximum eccentricity of the inner 
companion. The bottom left panel (c) is for the maximum  of
$\theta=\varpi_{\idm{b}}-\varpi_{\idm{c}}$ centered about $0^{\circ}$, the
bottom right panel (d) is for $\theta$ centered about $180^{\circ}$.  The
integration time is about $10^4$ periods of the outer planet. The thick curve
represents the planetary collision line given through
$a_{\idm{b}}(1+e_{\idm{b}}) = a_{\idm{c}}(1-e_{\idm{c}})$, where
$a_{\idm{b}}=1.49$~AU and $e_{\idm{c}}=0.31$ are fixed. Approximate
positions of the lowest order MMRs, relevant to the discussion,  are
labeled. The resolution of the plots is $200 \times 100$ points.
}
\label{fig:fig4}
\end{figure*} 

\begin{figure*}[th]
\centering
\includegraphics[]{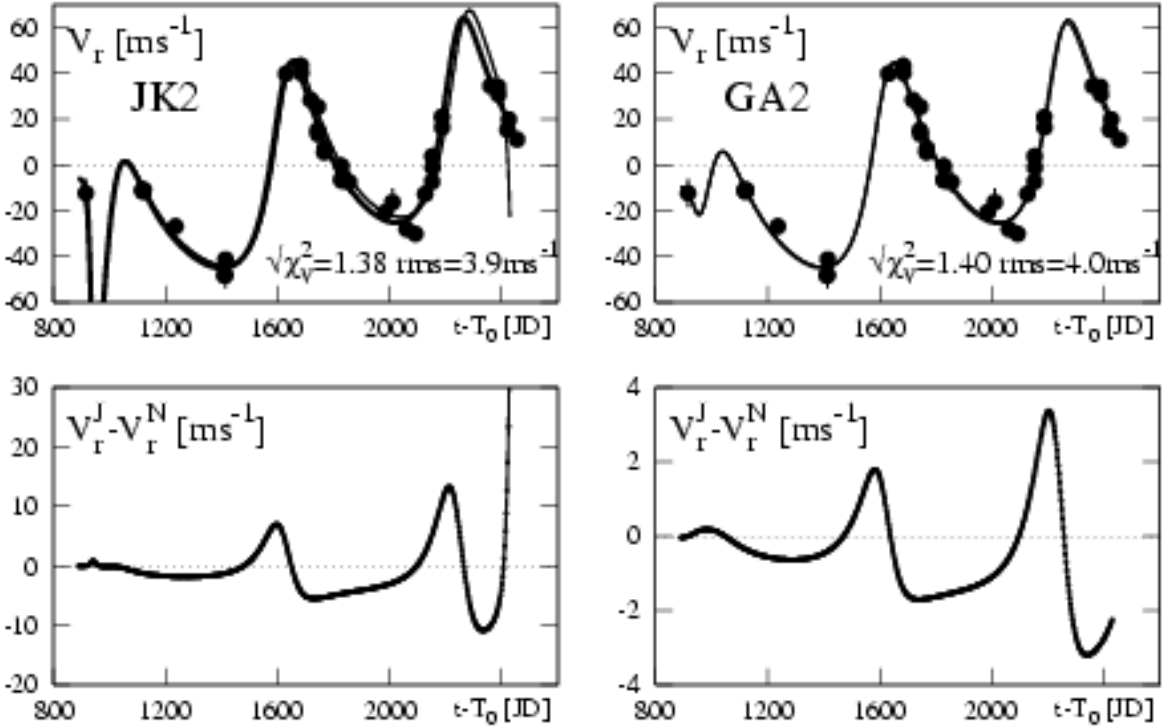}
\caption{
The top panels show a comparison of the synthetic RV curves, obtained from
the best double-Keplerian fits (JK2 and GA2, $V_{\idm{r}}^{\idm{J}}$), 
 and from the numerical
integration incorporating the  double-Keplerian solutions as the osculating
elements at the epoch of 
the first observation ($V_{\idm{r}}^{\idm{N}}$).  The left column is for the
best JK2 fit, the right column is for the best genetic fit GA2 (see
Table~\ref{tab:tab2}).  In order to make the comparison more transparent,
the bottom panels shows  the appropriate differences between the Keplerian
and Newtonian synthetic RV  signals. Time is given in Julian Days,
with $T_0=$JD2450000.
}
\label{fig:fig5}
\end{figure*}

\begin{figure}[th]
\centering
\hbox{
        \includegraphics[]{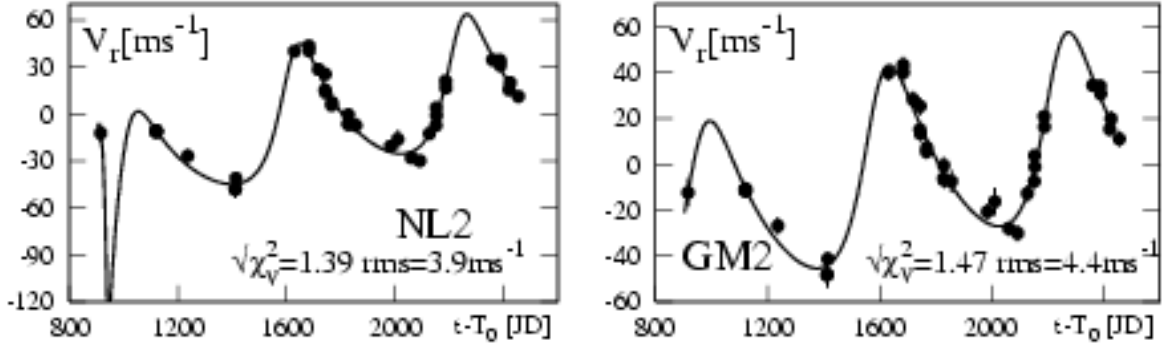}
}
\caption{
Synthetic RV curves for the best self-consistent fits  obtained by scanning
the parameters space with the Levenberg-Marquardt algorithm (NL2) and by the 
genetic  algorithm combined with the MEGNO signature (GM2).  The best fit
parameters are given in Tables~\ref{tab:tab2} and~\ref{tab:tab3}.
Time is given in Julian Days,
with $T_0=$JD2450000.
}
\label{fig:fig6}
\end{figure} 

\begin{figure*}[th]
\centering
\hbox{    
  \includegraphics[]{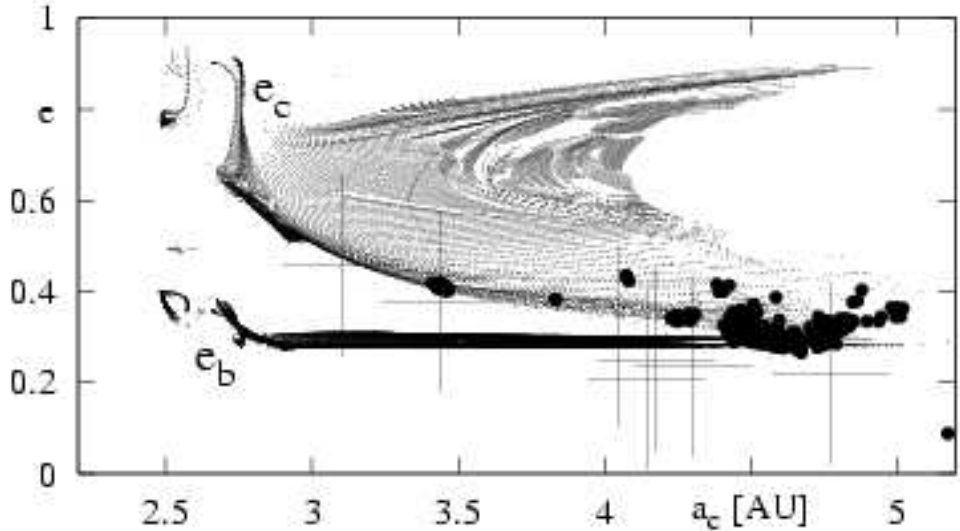}
     }
\caption{
A representation of the best self-consistent Newtonian fits obtained by the 
gradient method in the $(a_{\idm{c}},e_{\idm{c}})$-plane of astrocentric, 
osculating elements. The initial epoch is the Julian Date of the first
observation. The best double-Keplerian  fits have been used as the starting
points for the gradient method (see the text for more details). The dots
mark the fits which have $\Chi < 1.5$, $e_{\idm{b}}$, $e_{\idm{c}}$ stand for
the eccentricity of the inner and the outer companion, respectively. The
large, filled circles mark these $(a_{\idm{c}},e_{\idm{c}})$ of the fits that lead to a
stable, quasi-periodic evolution of the planetary system.  For these fits
$|\Ym-2| < 0.05$. Crosses  mark the initial parameters found by the
MEGNO-penalty fits (Table~\ref{tab:tab3}). 
}
\label{fig:fig7}
\end{figure*}

\begin{figure*}[th]
\centering
          \includegraphics[]{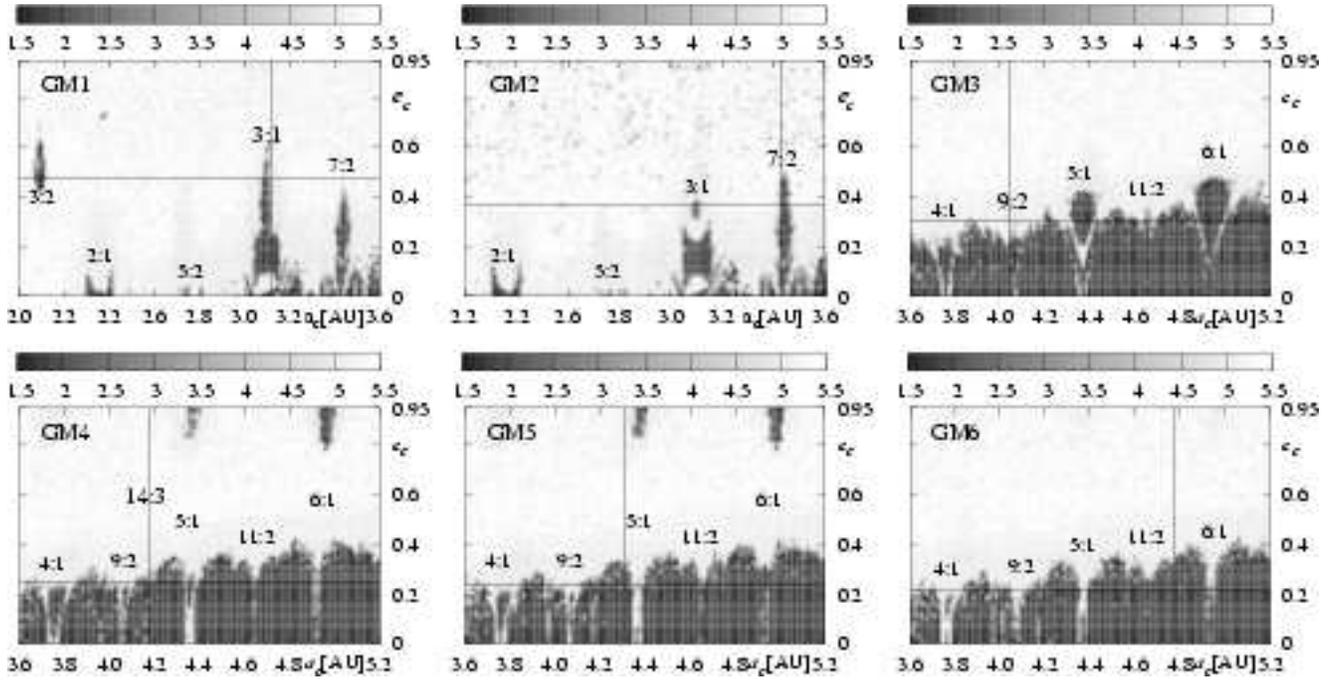}
\caption{
Stability maps  in the ($a_{\idm{c}},e_{\idm{c}})$-plane for the best genetic
fits GM1--GM6 (Table~\ref{tab:tab3}), 
obtained by applying the MEGNO penalty algorithm.
The fitted parameters,  given in Table~\ref{tab:tab3},  are marked by the
intersection of the two lines. The resolution of these scans is $160 \times
100$ data points.
}
\label{fig:fig8}
\end{figure*}

\begin{figure*}[th]
\centering
\hbox{    
  \includegraphics[]{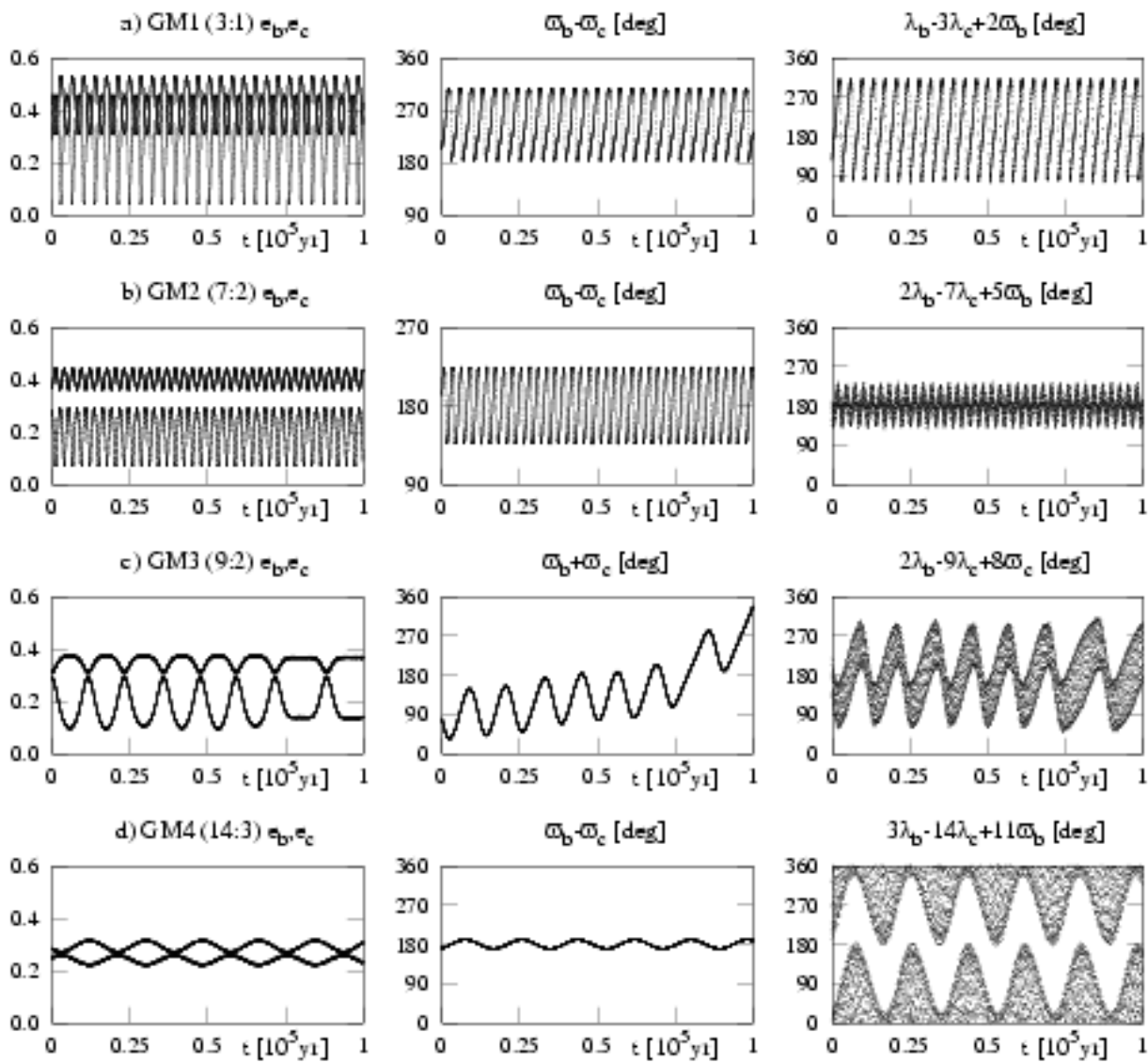}
        }
\caption{
Evolution of the orbital elements for the subsequent GM solutions 
(rows). For the GM1 solution $a_{\idm{c}}=3.1$AU. See also 
Table~\ref{tab:tab3}.}
\label{fig:fig9}
\end{figure*}

\begin{figure*}[th]
\centering
\hbox{    
  \includegraphics[]{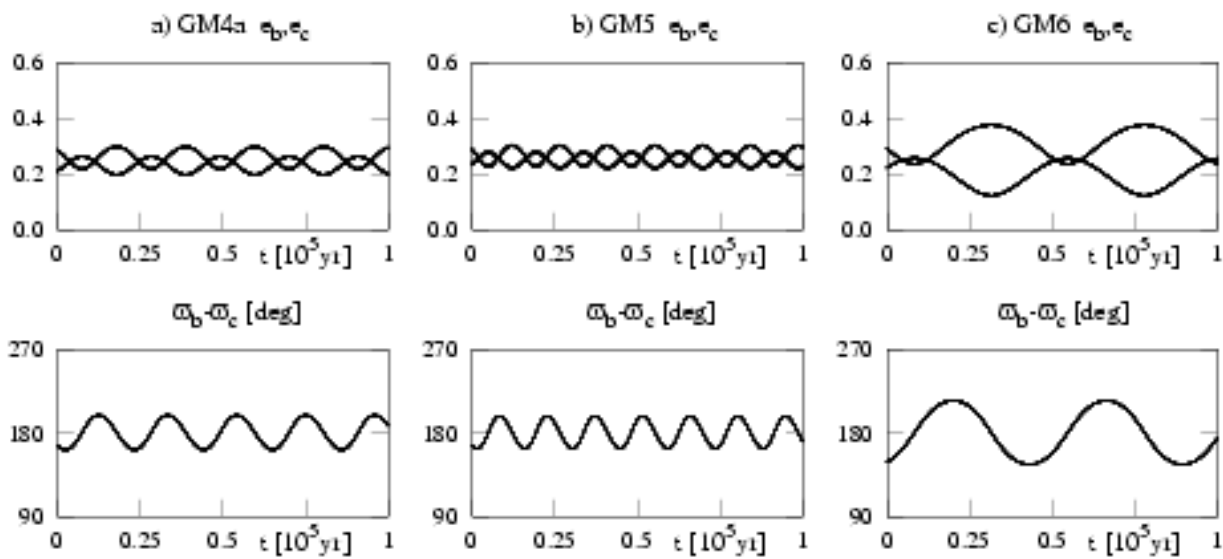}
     }
\caption{
Evolution of the orbital elements in the GM fits (columns). 
See also Table~\ref{tab:tab3}.
}
\label{fig:fig10}
\end{figure*}

\begin{figure}[th]
\centering
  \includegraphics[]{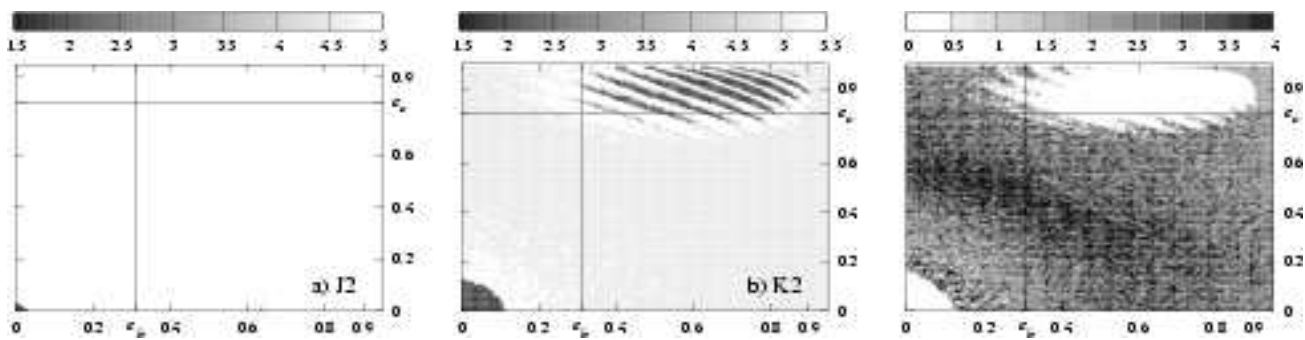}
\caption{
Stability maps in the $(e_{\idm{b}},e_{\idm{c}})$-plane: the left panel is
for the J2 fit from \cite{Jones2002}, the middle panel is for the
initial condition from \cite{Kiseleva2002} (the fit K2). The right
panel is for the fit K2 and it illustrates which systems are disrupted by a
collision or ejection of a planet.  Such  disrupted systems are marked with
different gray codes: 1 is for $e_{\idm{b}}>0.999$, 2 is for
$e_{\idm{c}}>0.999$, 3 is for $a_{\idm{b}}>10$~AU, 4 is for
$a_{\idm{b}}>10$~AU. Note the short timescale of the instabilities: the time
of integration is at most equal to about $10^4$ periods of the outer
companion.
The
resolution of the left plot is $100 \times 100$, for the two other plots it
is equal to $190 \times 190$ data points.
}
\label{fig:fig11}
\end{figure}

\begin{figure}[th]
\centering
\hbox{    
  \includegraphics[]{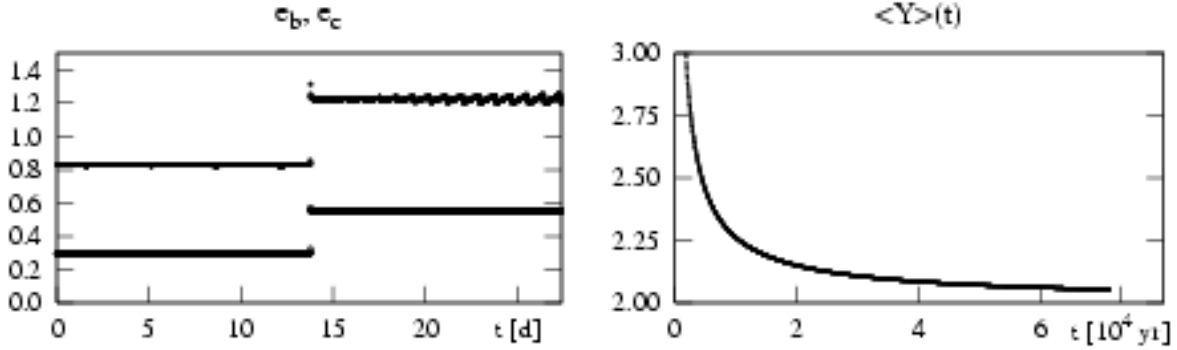}
     }
\caption{ 
An example of a misleading MEGNO convergence for a case of a self-disrupting
system. The left panel is for the initial condition J2 (Table~\ref{tab:tab1})
modified in such a way that $e_{\idm{b}}=0.29$ and $e_{\idm{c}}=0.83$. A
collision after about $14$~days disrupts the system. The right panel is for
MEGNO. The indicator converges to $\simeq 2.05$.
}
\label{fig:fig12}
\end{figure}

\begin{figure}[th]
\centering
\hbox{    
  \includegraphics[]{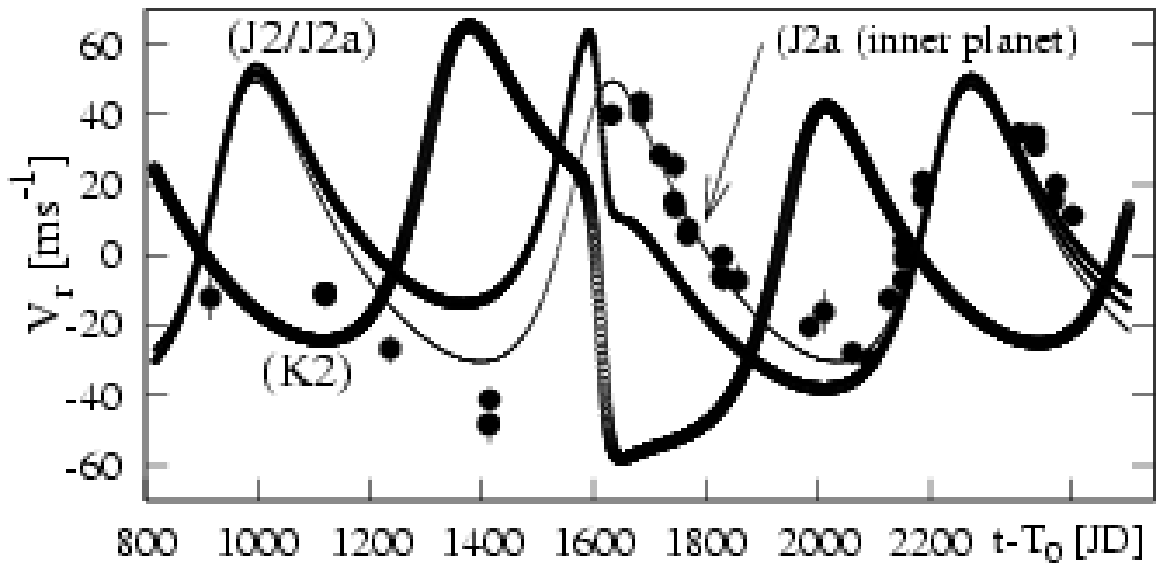}
     }
\caption{ 
Keplerian RV curves for the initial conditions given in \cite{Jones2002}
(J2/J2a) and \cite{Kiseleva2002} (K2). Small, filled circles are for the J2
and J2a fits. Open circles are for the K2 initial condition (see
Table~{\ref{tab:tab1}}). In all these cases, the RV offset is unspecified and
it has been set to $V_0=0~\ms$. Large, filled circles are for the RV
measurements published in \cite{Jones2003}.  Thin line is for the RV
corresponding to the signal of the inner planet only (its elements are given
by the J2a fit). Note that the RV signals have been calculated using the true
anomaly parameterization of the RV signal  (see the Appendix for
explanation). Time is given in Julian Days, with $T_0=$JD2450000.
}
\label{fig:fig13}
\end{figure}

\end{document}